%% file: main.tex
\documentclass[authoryear,3p, times,10pt,final]{elsarticle}

\usepackage{amsmath}

\usepackage[colorlinks=true,linkcolor=blue, pdfstartview=FitH]{hyperref}

\begin{document}

\begin{frontmatter}

\newtheorem{thm}{Theorem}
\newtheorem{lem}[thm]{Lemma}
\newdefinition{rmk}{Remark}
\newdefinition{note}{Note}
\newproof{pf}{Proof}
\newproof{pot}{Proof of Theorem \ref{thm2}}

\author[LGI2A]{Shahin Gelareh \corref{cor1}}

\cortext[cor1]{Corresponding author, shahin.gelareh@\{univ-artois.fr,gmail.com\}}


\address[LGI2A]{D\'epartement R\'eseaux et T\'el\'ecommunications, Universit\'{e} d'Artois, F-62400 B\'{e}thune, France}



\title{A note on 'Variable Neighborhood Search Based Algorithms for Crossdock Truck Assignment'}

\begin{abstract}
Some implementations of variable neighborhood search based algorithms were presented in  \emph{Cécilia Daquin, Hamid Allaoui, Gilles Goncalves and Tienté Hsu, Variable neighborhood search based algorithms for crossdock truck assignment, RAIRO-Oper. Res., 55 (2021) 2291-2323}. This work is based on model in \emph{Zhaowei Miao, Andrew Lim, Hong Ma, Truck dock assignment problem with operational time constraint within crossdocks, European Journal of Operational Research
 192 (1), 2009,  105-115 }(see \cite{MIAO2009105}) which has been proven to be incorrect. We reiterate and elaborate on the deficiencies in the latter and show that the authors in the former were already aware of the deficiencies in the latter and the proposed minor amendment does not overcome any of such deficiencies.

\end{abstract}
\end{frontmatter}

%
\section{Introduction} 

\cite{refId0} proposed implementation of a set of Variable Neighborhood Search (VNS) techniques to solve the problem proposed in \cite{MIAO2009105}. In \cite{MIAO2009105}, the authors proposed a mathematical formulation for truck dock assignment in a crossdock. The optimal solution of this model is then used as a reference point for  measuring the quality of their proposed genetic algorithm. \\

In a joint work with one of the co-authors in \cite{refId0}, we have shown in \cite{GELAREH20161144} and more elaborated in \cite{RG}, we have shown that this model has some serious issues and fails to deliver the expected outcome. More precisely we have shown by example that the model may deliver '\textit{optimal}' solutions that are more than 45 percent away from the real optimal solution. In short, the problem description and the proposed model do not match. \\

In \cite{refId0}, the authors introduce a minor coefficient in the cost function intending to correct the model itself. However, it is obvious that no modification in the objective can be a correction to the polyhedral structure. In the following, we re-iterate and re-emphasize on those issues.\\

\section{Mathematical Model}

According to \citep{MIAO2009105,Miao2006}: \emph{an over-constrained truck dock assignment problem with time window, operational time, and capacity constraint in a transshipment network through cross docks where the number of trucks exceeds the number
of docks available and the capacity of the cross dock is limited, and where the objective is to minimize the operational cost
of the cargo shipment and the number of unfulﬁlled shipments is studied}. \\
The objective function accounts for the total cost of 1) dock operations, 2) penalties for the unfulﬁlled shipments.\\

The following parameters and variables are introduced therein:\\

\textit{Parameters:}\
\begin{description}\itemsep1pt \parskip0pt \parsep0pt
\item $N$:\, set of trucks arriving at and/or departing from the cross dock
\item $M$:\, set of docks available in the cross dock
\item $n$:\, total number of trucks, that is $|N|$ ($|N|$ denotes cardinality of set $N$)
\item $m$:\, total number of docks, that is $|M|$
\item $a_i$:\, arrival time of truck $i$ ($1\leq i \leq n$)
\item $di$:\, departure time of truck $i$ ($1\leq i \leq n$)
\item $t_{kl}$:\, operational time for pallets from dock $k$ to dock $l$ ($1\leq k,l \leq m$)
\item $f_{ij}$:\, number of pallets transferring from truck $i$ to truck $j$ ($1\leq i,j \leq n$)
\item $c_{kl}$:\, operational cost per unit time from dock $k$ to dock $l$ ($1\leq k,l \leq m$)
\item $p_{ij}$:\,  penalty cost per unit cargo from truck $i$ to truck $j$ ($1\leq i,j \leq n$)
\item $C$:\, capacity of cross dock, i.e. the maximum number of cargos the cross dock can hold at a time
\item $\hat{x}_{ij}$:\, 1 iff truck $i$ departs no later than truck $j$ arrives; 0 otherwise.
\end{description}

\begin{note}[\cite{MIAO2009105}]\label{lbl:note1}
It has been also assumed that:
 \begin{itemize}[-]
 \item $f_{ij}\geq 0$ iff $d_j\geq a_i, \forall i,j\neq i$, otherwise $f_{ij}= 0$ meaning that truck $i$ will transfer some cargo to truck $j$ iff truck $j$ departs no earlier than truck $i$ arrives;
 \item  $a_i< d_i (1\leq i \leq n)$ which means for each truck, the arrival time should strictly smaller than its departure time;
  \item $n>m$ which satisfies the over-constrained condition;
  \item sort all the $a_i$ and $d_i$ in an increasing order, and let $t_r~ (r = 1,2,. . .,2n)$ correspond to these $2n$ numbers such that
$t_1 \leq t_2 \leq \dots \leq t_{2n}$. Using this notation, we can easily formulate the set of capacity constraints later.
  \end{itemize}

\end{note}

\textit{Variables:}\itemsep1pt \parskip0pt \parsep0pt
\begin{description}\itemsep1pt \parskip0pt \parsep0pt
\item $y_{ik}$:\,1 if truck $i$ is assigned to dock $k$; 0, otherwise.
\item $z_{ijkl}$:\,1 if truck $i$ is assigned to dock $k$ and truck $j$ is assigned to dock $l$; 0 otherwise.
\end{description}
\subsection{Mathematical model}
We call this problem CROSS-DOCK:\\

[CROSS-DOCK]\\
\begin{align}
min & \sum_{k=1}^m \sum_{l=1}^m \sum_{i=1}^n \sum_{j=1}^n c_{kl}t_{kl}z_{ijkl} + \sum_{i=1}^{n} \left(\sum_{j=1}^{n} p_{ij}f_{ij}\left(1- \sum_{k=1}^m\sum_{l=1}^m z_{ijkl}\right) \right) \label{orig::obj}\\
s.t. & \nonumber\\
    &   \sum_{k=1}^m y_{ik} \leq 1,     &   \forall i \label{orig::eq1}\\
    &   z_{ijkl}\leq y_{ik} ,     &   \forall i,j, k,l :j\neq i \label{orig::eq2}\\
    &   z_{ijkl}\leq y_{jl} ,     &   \forall i,j, k,l  :j\neq i \label{orig::eq3}\\
    &   y_{ik}+y_{jl} -1 \leq z_{ijkl}  &   \forall i,j,k,l  :j\neq i \label{orig::eq4}\\
    &   \hat{x}_{ij}+\hat{x}_{ji} \geq z_{ijkk}  &   \forall i,j,k,l :j\neq i \label{orig::eq5}\\
    &   \sum_{k=1}^m \sum_{l=1}^m \sum_j^n\sum_{i\in \{i:a_i\leq t_r\}} f_{ij}z_{ijkl} - \sum_{k=1}^m \sum_{l=1}^m \sum_i^n \sum_{j\in \{j:d_j\leq t_r\}} f_{ij}z_{ijkl} \leq C   &   \forall r\in \{1,2,\dots,2n\}  \label{orig::eq6}\\
    &       f_{ij}z_{ijkl}(d_j-a_i-t_{kl})\geq 0        &       \forall i,j,k,l :j\neq i \label{orig::eq7}\\
    &   y_{ik}\in\{0,1\}, z_{ijkl}\in \{0,1\}    \label{orig::eq8}
\end{align}

The first part in \eqref{orig::obj} accounts for the cost of transport, and the second part calculates the penalty for the unfulfilled transports. \eqref{orig::eq1} ensures that a truck can be assigned to no more than one dock. Consider constraints \eqref{orig::eq7} before the others. \eqref{orig::eq7} indicates that, if truck $i$ uses dock $k$ and truck $j$ uses dock $l$ then the variable $z_{ijkl}$ \emph{may} take 1, only if the arrival of $i$ plus the transfer time from $k$ to $l$ is not later than the departure of $j$. That means, there is a sense of \emph{direction of flow} associated to each variable $z_{ijkl}$ which is not covered in the definition of this variable in \citep{MIAO2009105}. Therefore, to this end, we associate such flow direction to the variable $z_{ijkl}$.\\

\eqref{orig::eq2}-\eqref{orig::eq3} indicate that for a pallet transfer between truck $i$ and truck $j$, $i$ can use dock $k$ and $j$ can use dock $l$, if $i$ is allocated to $k$ and $j$ is allocated to $l$. Constraints \eqref{orig::eq4} imply that if the truck $i$ is assigned to dock $k$ and truck $j$ to  dock $l$, a bidirectional transfer between the two trucks ---from $i$ to $j$ represented by $z_{ijkl}$ and from $j$ to $i$ represented by $z_{jilk}$--- must take place. 
If the transfer does not take place, then \emph{not} both trucks can be docked. Moreover, if transfer from $i$ to $j$ does not take place, from $j$ to $i$ must not take place either. I.e. $z_{ijkl}=1$ iff $z_{jilk}=1$ and $z_{ijkl}=0$ iff $z_{jilk}=0$. Constraints \eqref{orig::eq5} ensure that truck $i$ and truck $j$ can use the same dock for realizing the transfer of pallets from $i$ to $j$, only if their time windows do not intersect ---i.e., $i$ leaves no later than $j$ arrives. Constraints \eqref{orig::eq6} guarantee that at every event time (arrival and/or departure of a truck), the capacity of cross dock is respected.\\
For a given $i,j,k,l:j\neq i$, the corresponding constraint in \eqref{orig::eq7} exists only if $f_{ij}$ and $(d_j-a_i-t_{kl})$ are nonzero, otherwise $f_{ij}$ or $(d_j-a_i-t_{kl})$ would void the constraint. As a consequence,  this would be $z_{jilk}$ which influences the value that $z_{ijkl}$ must take ---as the left-hand side of \eqref{orig::eq4} is the same for $z_{ijkl}$ and $z_{jilk}$.\\

\section{Issues in the model}\label{sec::issue}
The model possesses several issues some of which we could identify are listed in the sequel:
\subsection{Constraints \eqref{orig::eq4}  eliminate correct solutions of the problem}
Case 1: Given $i,j\neq i$ such that $f_{ij}=0$, according to Note \ref{lbl:note1}, if $j$ would have already left before $i$ arrives then $f_{ij}=0$ and $i$ would not be able to deliver to $j$. But, if $f_{ji}\neq 0$ and a sufficient capacity on the cross dock would have been available, truck $j$ might have already dropped off its cargo on the cross dock buffer ---before it departs. This cargo  can be later on transferred to $i$, once $i$ arrives. Having $f_{ij}=0$, voids the corresponding constraint in \eqref{orig::eq7} and this constraint does not make any decision on $z_{ijkl}$. If for some $k,l$, $z_{jilk}=1$, as it is possible, then constraint  \eqref{orig::eq4} forces $z_{ijkl}$ to take 1, too, for a zero-size load of pallets from $i$ to $j$.\\

When $f_{ij}=0$ by Note \ref{lbl:note1} and $z_{ijkl}=1$ (due to equality with $z_{jilk}$), we have $p_{ij}f_{ij}\left(1- \sum_{k=1}^m\sum_{l=1}^m z_{ijkl}\right)$ equal to zero because no penalty is going to be paid for a flow that does not exist, i.e. $f_{ij}=0$. However, still the first term in the objective function (i.e. for the same $i,j$, $\sum_{k=1}^m \sum_{l=1}^m  c_{kl}t_{kl}z_{ijkl} $) is contributing in the objective values by forcing to pay transport cost for pallets, which do not exist. That means, for $f_{ij}=0$ we still have to pay for the transfer cost.\\

This is a direct consequence of  tying up the destiny of $z_{ijkl}$ and $z_{jilk}$ to each other using \eqref{orig::eq4}.

Case 2: Given $i,j,k,l:j\neq i$, suppose that $(d_j-a_i)\geq 0$  and $(d_j-a_i-t_{kl})\leq 0$ such that $j$ does not leave before $i$ arrives, but there is no sufficient time to make the transfer between dock $k$ and dock $l$. However, if $j$ has left something on the buffer before it leaves, $i$ which arrives later can still take it and $z_{jilk}$ can take 1. This does not fall into the category of Note \ref{lbl:note1} and $f_{ij}$ is not necessarily 0 (might be strictly positive). Therefore, corresponding constraint in \eqref{orig::eq7} exists and forces $z_{ijkl}=0$. On the other hand, \eqref{orig::eq4} forces $z_{jilk}$ to take 0 and avoids it to take 1. \

In such a case, having $z_{jilk}=1$ in a feasible solution of the real-life problem is possible (assuming a sufficient capacity), but it is eliminated by this model.\

Therefore, forcing $z_{jilk}$ and $z_{ijkl}$ to take the same values (i.e. having \eqref{orig::eq4} in the model) is \emph{incorrect} and those constraints imposing such conditions must be removed from CROSS-DOCK.

\subsection{Constraints \eqref{orig::eq5} }
Because a variable $z_{ijkl}$ implies a sense of direction of flow, we make the following re-definition:
\begin{description}
\item $z_{ijkl}$:\,1 if pallets of truck $i$ which is assigned to dock $k$ are transferred to the truck $j$ which is assigned to dock $l$; 0 otherwise.
\end{description}

There are two different cases allowing two distinct trucks $i$ and $j$ use the same dock $k$: 1) truck $i$ drops its pallets and leaves dock $k$ before $j$ arrives at dock $k$, 2) truck $j$ drops its pallets and leaves dock $k$ before truck $i$ arrives at dock $k$. Therefore, $z_{ijkk}$ only depends on whether $i$ leaves before $j$ arrives or not. Whether $j$ departs before $i$ arrives or not is \emph{not} directly related to $z_{ijkk}$. Because if $j$ depart before $i$ arrives (i.e. $\hat{x}_{ij}=0$ and $\hat{x}_{ji}=1$), still there is no possibility of transferring from $i$ to $j$, but the constraint turns to $z_{ijkk}\leq \hat{x}_{ij} + \hat{x}_{ji} =1$ which is of no effect unless it causes numerical deficiencies. While $z_{ijkk}\leq \hat{x}_{ij} = 0$, clearly sets the variable to the correct value. Consequently, constraints \eqref{orig::eq5} should be replaced by:
\begin{align}
&   \hat{x}_{ij} \geq z_{ijkk}  &   \forall i,j\neq i,k,l
\end{align}

\section{Example}\label{sec::example}
\input{example}

The \emph{claim} is that this model does not correctly represent the problem. In other words, there are other feasible solutions with better objective values to the problem which are eliminated by this model (in particular, constraints \eqref{orig::eq4}).\

We remove constraint \eqref{orig::eq4} and rectify \eqref{orig::eq5} and call the model R-CROSS-DOCK:\

[R-CROSS-DOCK]\
\begin{align}
min & \sum_{k=1}^m \sum_{l=1}^m \sum_{i=1}^n \sum_{j=1}^n c_{kl}t_{kl}z_{ijk} + \sum_{i=1}^{n} \left(\sum_{j=1}^{n} p_{ij}f_{ij}\left(1- \sum_{k=1}^m\sum_{l=1}^m z_{ijkl}\right) \right) \label{revised::obj}\\
s.t. & \nonumber\\
    &   \sum_{k=1}^m y_{ik} \leq 1,     &   \forall i \label{revised::eq1}\\
    &   z_{ijkl}\leq y_{ik} ,     &   \forall i,j, k,l \label{revised::eq2}\\
    &   z_{ijkl}\leq y_{jl} ,     &   \forall i,j, k,l \label{revised::eq3}\\
    &   y_{ik}+y_{jk}  \leq 1+ \hat{x}_{ij}+\hat{x}_{ji}  &   \forall i,j,k \label{revised::eq4}\\
    &   z_{ijkk} \leq \hat{x}_{ij}    &   \forall i,j\neq i,k,l \label{revised::eq5}\\
    &   \sum_{k=1}^m \sum_{l=1}^m \sum_j^n\sum_{i\in \{i:a_i\leq t_r\}} f_{ij}z_{ijkl} + \sum_{k=1}^m \sum_{l=1}^m \sum_i^n \sum_{j\in \{j:d_j\leq t_r\}} f_{ij}z_{ijkl} \leq C   &   \forall r\in \{1,2,\dots,2n\}  \label{revised::eq6}\\
    &   z_{ijkl} = 0       &       \forall i,j,k,l:j\neq i, (d_j-a_i-t_{kl})\leq 0   \label{revised::eq7}\\
    &   y_{ik}\in\{0,1\}, z_{ijkl}\in \{0,1\}    \label{revised::eq8}
\end{align}
The new constraints \eqref{revised::eq4} ensure that if the arrival/departure time windows of two trucks $i$ and $j$ overlap ($\hat{x}_{ij}=\hat{x}_{ji}=0$), either of them can be docked at a dock $k$ (not both).
The optimal solution, $s'^*$, of this problem using the model CROSS-DOCK has an optimal objective value of  11 and $z_{1311} = 1,~
z_{1412} = 1,z_{2321} = 1,z_{2422} = 1,z_{4321} = 1,y_{11} = 1,y_{22} = 1,y_{31} = 1,y_{42} = 1$ while $\hat{x}_{13} = 1,\hat{x}_{14} = 1,\hat{x}_{23} = 1,\hat{x}_{24} = 1$ and the relative gap between the two solutions is 45.45 \%.

\subsection{Why $s'^*$ is not feasible in $s^*$}
A diagnosis of  infeasibility has reported that an \emph{Irreducibly Inconsistent Set (IIS)} of constraints is consisted of  $y_{11} + y_{22} <= 1+z_{1212}$ as the cause of infeasibility.\

In $s'^*$, we have $y_{11} = 1,~y_{22} = 1$. By substituting in constraint \eqref{orig::eq5}, we have $1+1 = y_{11} + y_{22} <= 1+z_{1212}$ ($1+1 = y_{11} + y_{22} <= 1+z_{2121}$) which forces $z_{2121}=1$. However, constraint \eqref{orig::eq7} has already suggested $z_{1212}=0$ (and $z_{2121}=0$) as $d_j-a_i-t_{kl}<0$ (and $d_i-a_j-t_{lk}<0$) which is a contradiction.\

\section{Summary and conclusion}
We carefully devised an instance of the problem and showed that the model in \cite{MIAO2009105} has some fundamental polyhedral deficiencies and does not correspond to the problem definition they proposed therein. This model does not necessarily produce an optimal solution and cannot be used as a reference point to measure the quality of heuristic solutions. The amendment proposed in \cite{refId0} does not touch the polytope and therefore, all the structural issues remain intact.


\bibliographystyle{apalike}
\bibliography{sample}


\end{document}

%% file: example.tex
Let $m$ number of docks be 6 and $n$ number of trucks be 9. Moreover, let:
$\tiny
c=t=\left(
   \begin{array}{cccccc}
     0& 1& 2& 3& 4& 5\\
    1& 0& 1& 4& 3& 4 \\
    2& 1& 0& 5& 4& 3 \\
    3& 4& 5& 0& 1& 2 \\
    4& 3& 4& 1& 0& 1 \\
    5& 4& 3& 2& 1& 0 \\
   \end{array}
 \right)
 $,
 $\tiny
p=f=10\times\left(
   \begin{array}{ccccccccc}
 19&	13&	19&	14&	12&	13&	16&	12&	14\\
19&	18&	16&	17&	18&	14&	12&	14&	16\\
18&	17&	15&	20&	13&	13&	17&	15&	17\\
10&	11&	10&	10&	19&	19&	16&	20&	14\\
19&	20&	19&	19&	12&	18&	20&	10&	15\\
20&	17&	12&	15&	14&	20&	20&	17&	10\\
18&	14&	13&	10&	19&	20&	15&	19&	18\\
15&	11&	20&	20&	14&	12&	18&	13&	10\\
17&	10&	12&	10&	18&	13&	18&	20&	20\\
  \end{array}
 \right)
 $,
 $\tiny
 a=\left(
   \begin{array}{c}
         15.42\\
        15.50\\
        17.00\\
        16.52\\
        16.41\\
        16.08\\
        16.52\\
        16.28\\
        16.29\\
  \end{array}
 \right)
 $
 and
 $\tiny
 d=\left(
   \begin{array}{c}
   16.41\\
    16.41\\
    18.00\\
    17.57\\
    17.46\\
    17.10\\
    18.05\\
    17.34\\
    17.42\\
  \end{array}
 \right)
 $\\

 The optimal solution, $s^*$, of this problem using the model CROSS-DOCK has an optimal objective value of  316951.0 and $z_{1312} = 1,~z_{1713} = 1,~z_{3121} = 1,~z_{3723} = 1,~z_{7131} = 1,~z_{7332} = 1,~y_{11} = 1,~y_{32} = 1,~y_{73} = 1$ while $\hat{x}_{13} = 1\hat{x}_{14} = 1,~\hat{x}_{15} = 1,~
\hat{x}_{17} = 1,~\hat{x}_{23} = 1,~\hat{x}_{24} = 1,~\hat{x}_{25} = 1,$ and ~$\hat{x}_{27} = 1$\\